\documentclass[useAMS,usenatbib]{mn2e}
\usepackage{subfigure}
\usepackage{graphicx}
\newcommand{\Sec}{Section~}

\title[Torque bistability]{Torque bistability in the interaction
between a neutron star magnetosphere and a thin accretion disc}
\author[J. T. Locsei and A. Melatos]{J. T. Locsei$^{1}$\thanks{E-mail:
tlocsei@physics.unimelb.edu.au } and
A. Melatos$^{1}$\\
$^{1}$School of Physics, University of Melbourne, Parkville VIC 3010,
Australia}

\begin{document}

\date{Accepted 1988 December 15. Received 1988 December 14; in original form 1988 October 11}

\pagerange{\pageref{firstpage}--\pageref{lastpage}} \pubyear{2002}

\maketitle

\label{firstpage}

\begin{abstract}
We present a time-dependent model of the interaction between a neutron
star magnetosphere and a thin (Shakura-Sunyaev) accretion disk, where
the extent of the magnetosphere is determined by balancing outward
diffusion and inward advection of the stellar magnetic field at the
inner edge of the disc. The nature of the equilibria available to the
system is governed by the magnetic Prandtl number ${\rm Pm}$ and the
ratio $\xi$ of the corotation radius to the Alfv\'{e}n radius. For $\xi
\ga {\rm Pm}^{0.3}$, the system can occupy one of two stable states
where the torques are of opposite signs. If the star is spinning up
initially, in the absence of extraneous perturbations, $\xi$ decreases
until the spin-up equilibrium vanishes, the star subsequently spins
down, and the torque asymptotes to zero. Vortex-in-cell simulations of
the Kelvin-Helmholtz instability suggest that the transport speed
across the mixing layer between the disc and magnetosphere is less than
the shear speed when the layer is thin, unlike in previous models.
\end{abstract}


\section{Introduction}

A complete \emph{time dependent} theory of mass and angular momentum
transfer onto a neutron star accreting via a thin (Shakura-Sunyaev)
disc needs to explain the stochastic fluctuations observed in the spin
frequencies and X-ray luminosities of such systems
\citep{Bildsten.1997,Baykal.1993}, as well as regular phenomena like
the negative (spin-down) torque on accreting millisecond pulsars
\citep{Galloway.2002,Chakrabarty.2003}, quasi-periodic oscillations
(QPOs) in low-mass X-ray binaries (LMXBs) \citep{vanderKlis.2001}, and
torque reversals \citep{Nelson.1997}. Torque reversals, in which some
X-ray pulsars alternate between sustained episodes of spin-up/down,
present a particular challenge to theories of accretion because the
transitions between spin-up/down episodes are sudden ($\sim 10$\% of
the duration of the episodes themselves).

In the standard \citep{Ghosh.1979a, Ghosh.1979b} picture of neutron
star accretion, a thin disc is threaded by the stellar magnetic field.
The torque on the star is the sum of the material torque $N_{\rm mat}$
carried by material falling onto the star and the torque $N_B$ due to
magnetic stresses. The material torque $N_{\rm mat}$ always acts to
spin up the star, while the magnetic torque $N_B$ may either spin up or
spin down the star. The Ghosh-Lamb picture has been extended in several
directions to accommodate time-dependent effects. \citet{Lovelace.1995}
proposed that the magnetosphere could alternate between configurations
with and without outflows. \citet{Li.1998b} suggested that the maximum
radius where the stellar magnetic field threads the disc is constrained
to small and large values, with intermediate values excluded by an
unstable feedback mechanism as the magnetospheric configuration alters
the disc resistivity. \citet{Torkelsson.1998} considered the
interaction between a disc dynamo and the stellar magnetic field.
\citet{Kerkwijk.1998} considered the possibility that the inner disc
flips over, due to warping, to form a retrograde disc. \citet{Yi.1999}
suggested that the accretion flow changes abruptly from a geometrically
thin and cool Keplerian flow to a geometrically thick and hot
sub-Keplerian flow, accompanied by a change from spin-up to spin-down.
\citet{Spruit.1993} (ST93 hereafter) showed that the disc can become
unstable, with mass accreted in bursts, based on the premise that
accretion is enhanced by the Kelvin-Helmholtz instability (KHI) but
inhibited when the magnetosphere rotates faster than the disc
\citep{Illarionov.1975}. They compared the bursts predicted by their
model to the type II bursts observed from the Rapid Burster \mbox{MXB
1730--355}.

In this paper, we generalize the accretion model of ST93 to include
torque feedback onto the star and diffusive penetration of the stellar
magnetic field into the inner edge of the accretion disc, balanced by
inward advection. On the basis of preliminary numerical simulations, we
also neglect enhancement of accretion by the KHI. We present the new
model in \Sec\ref{sec:newmodel}. In \Sec\ref{sec:bistability}, we show
that the disc-magnetosphere system has two stable torque states, for
certain combinations of the magnetic Prandtl number ${\rm Pm}$ and the
ratio $\xi$ of the corotation radius to the Alfv\'{e}n radius. If the
star is initially spinning up, in the absence of extraneous
perturbations, the spin-up equilibrium eventually vanishes and the star
subsequently spins down. In \Sec\ref{sec:flaw_in_framework}, we show
that the viscous torque in the mixing layer, omitted in the original
ST93 model, is significant; when it is included, the region of
parameter space for which the system has two stable torque states is
reduced. In \Sec\ref{sec:mhd_instabilities}, we present the results of
preliminary numerical simulations which indicate that the transport of
material across the disc-magnetosphere mixing layer due to the KHI is
suppressed when the mixing layer is thin. In \Sec\ref{sec:conclusions},
we discuss the implications of the generalized model for the torque
reversals observed in accreting X-ray pulsars.

\section[]{Disc-magnetosphere coupling}
\label{sec:newmodel}

\subsection{Magnetospheric radius}
\label{subsec:magnetospheric_radius}

We use cylindrical polar coordinates $(r, \phi, z)$ throughout and
assume axisymmetry. The magnetospheric radius $r_{\rm m1}$ of the
system is defined to include all material that corotates with the star
(ST93). [Such a radius may not exist if the magnetic field lines are
swept back all the way to the stellar surface; \citet{Ghosh.1977}].
Outside $r_{\rm m1}$, there is a thin, annular mixing layer $r_{\rm m1}
< r < r_{\rm m2}$ within which the angular velocity of infalling
material is brought from the Keplerian value $\Omega_{\rm K}(r_{\rm
m2})$ to the stellar velocity $\Omega_*$ by magnetic torques. Once the
material enters the mixing layer, it is transported on to the star on a
time-scale much shorter than $r_{\rm m1}/\dot{r}_{\rm m1}$. The rate at
which mass flows inward through a surface at a fixed radius $r$ in the
disc is
\begin{equation} \label{eq:mdot}
   \dot{M}(r,t) = -2 \pi r v_r(r,t) \Sigma (r, t),
\end{equation}
where $\Sigma$ is the surface mass density, and $v_r < 0$ is the radial
drift velocity. In the steady state, $\dot{M}$ is constant, independent
of $r$ (by continuity), and equals the accretion rate onto the star.
Out of the steady state, the accretion rate onto the star is
\begin{equation} \label{eq:mdotprime}
   \dot{M}'(t) = \dot{M}(r_{\rm m2}, t)
   + 2 \pi r_{\rm m2} \dot{r}_{\rm m2} \Sigma (r_{\rm m2}, t),
\end{equation}
where the second term in (\ref{eq:mdotprime}) represents mass swept up
as $r_{\rm m2}$ changes.

We place an upper bound on the steady state value of $r_{\rm m1}$ by
noting that the torque required to maintain corotation in the region
$r<r_{\rm m1}$ cannot exceed that provided by the magnetic shear
stress, yielding (ST93)
\begin{equation} \label{eq:rm1max}
   r_{\rm m1,max} = \frac{\dot{M} \Omega_*}
      {\pi \left| S(r_{\rm m1,max}) \right|},
\end{equation}
where $S(r)$, the magnetic stress at radius $r$, at the disc surface
($z=h$), is given by \citep{Shapiro.1983}
\begin{equation} \label{eq:mag_stress_BzBphi}
   \left| S(r) \right| =
      (2\pi)^{-1} \left| B_z(r) B_\phi(r) \right|_{z=h}.
\end{equation}
The toroidal field, generated by differential rotation, is limited to
$B_\phi \la B_z$, because as $B_\phi$ builds up, the magnetic pressure
inflates the field lines outward until they open up, destroying the
magnetic link between the star and the disc \citep{Aly.1985,Aly.1988}.
Hence, we have
\begin{equation} \label{eq:magstress_spruit}
   \left| S(r) \right| = \eta B^2(r) \, (4 \pi)^{-1}
   \approx \eta \mu^2 r^{-6} (4\pi)^{-1},
\end{equation}
where $B(r)$ is the strength of the dipole field in the absence of a
disc, $\mu$ is the stellar dipole moment, and $\eta \sim 1$
encapsulates our ignorance of the true (time dependent) magnetic field
modified by the disc.

The system does not necessarily adjust to $r_{\rm m1}=r_{\rm m1,max}$.
The value of $r_{\rm m1}$ depends on how rapidly the magnetic field
(which induces corotation) penetrates the disc by Ohmic diffusion. Let
$d_B=r_{\rm m2}-r_{\rm m1}$ be the depth to which the magnetic field
penetrates the disc beyond $r_{\rm m1}$, and let $\Delta j$ be the
change in specific angular momentum across the mixing layer, given by
\begin{equation} \label{eq:delta_ell}
   \Delta j =
   r_{\rm m1}^2 \Omega_* - r_{\rm m2}^2 \Omega_{\rm K}(r_{\rm m2}).
\end{equation}
Define $d_{\rm c}$ to be the critical depth to which the field would
need to penetrate in order to change the specific angular momentum of
inflowing material by $\Delta j$, obtained by setting $\dot{M} \Delta j
= -2 \pi r_{\rm m1}^2 d_{\rm c} S(r_{\rm m1})$. One then finds
\begin{equation} \label{eq:dcfirst}
   d_{\rm c} = \frac{\dot{M}(r_{\rm m2}) \Omega_*}{2 \pi S(r_{\rm m1})}
      \left[
      \left( \frac{r_{\rm m2}}{r_{\rm m1}} \right) ^2
      \left( \frac{r_{\rm m2}}{r_{\rm c}} \right) ^ {-3/2} - 1
      \right],
\end{equation}
where the corotation radius is given by $r_{\rm c} = ({\rm G}M /
\Omega_*^2)^{1/3}$, and our sign convention is such that a positive
magnetic stress $S$ subtracts angular momentum from the disc, as in
ST93. Since the magnetic stress acts in the direction to bring material
crossing the mixing layer into corotation, one has ${\rm sgn}(S) = {\rm
sgn}(r_{\rm c} - r_{\rm m1})$.

In equilibrium, one has $d_B = d_{\rm c}$. If the equilibrium is
perturbed by increasing $\dot{M}$, $d_{\rm c}$ increases, inflowing
material cannot be brought into corotation as it travels from $r_{\rm
m2}$ to $r_{\rm m1}$, and $r_{\rm m1}$ must decrease by an amount
$\approx d_{\rm c} - d_B$ on the local radial drift time $\approx
\left| v_r(r_{\rm m2}) \right| / d_B$, giving
\begin{equation} \label{eq:rm1dot}
   \dot{r}_{\rm m1} \approx
   \frac{d_B - d_{\rm c}}{d_B} \left| v_r(r_{\rm m2}) \right|.
\end{equation}
Similarly, if the equilibrium is perturbed by reducing $\dot{M}$,
$d_{\rm c}$ decreases, inflowing material is brought into corotation
before travelling the full distance from $r_{\rm m2}$ to $r_{\rm m1}$,
and $r_{\rm m1}$ must increase by an amount $\approx d_B - d_{\rm c}$
in a time $\approx \left| v_r(r_{\rm m2}) \right| / d_B$, as in
(\ref{eq:rm1dot}).

\subsection{Mixing layer thickness}
\label{subsec:blthickness}

The mixing layer thickness $d_B$ is determined by a competition between
outward diffusion and inward advection. Let $v_B$ be the speed at which
the magnetic field penetrates into the disc, as measured in a frame
where $v_r = 0$. One finds $v_B \approx \nu_B / d_B$, as for any
diffusive process, where $\nu_B$ is the magnetic diffusivity of the
disc material. In the observers frame, where the magnetospheric radius
is changing and matter flows radially inward, we have $v_B =
\dot{r}_{\rm m2} + \left| v_r(r_{\rm m2}) \right|$, and hence
\begin{eqnarray}  \label{eq:dBdot}
   \dot{r}_{\rm m2} = \frac{\nu_B}{d_B}
      - \left| v_r(r_{\rm m2}) \right|.
\end{eqnarray}
It might appear, from (\ref{eq:dBdot}), that we are erring by modelling
a diffusive process as an advective one. This is not so, as is clear
from the special case $\dot{r}_{\rm m1} = 0$, $v_r(r_{\rm m2}) = 0$,
where (\ref{eq:dBdot}) reduces to $\dot{d}_{\rm B} = \nu_B / d_B$ and
hence $d_B = (2 \nu_B t)^{1/2}$ for $d_B(t=0) = 0$, as expected for
diffusion.

Comparing our model with ST93, we note the following. ST93 set $r_{\rm
m1}$ to the largest value for which the corotation of material within
$r_{\rm m1}$ can be maintained by magnetic stresses. Subsequently, they
calculate the mixing layer thickness by requiring that disc material be
brought into corotation by magnetic stresses before reaching $r_{\rm
m1}$. In our model, the mixing layer thickness is set by the
advection-diffusion balance, and $r_{\rm m1}$ grows or shrinks
depending on whether or not material can be brought into corotation as
it crosses the mixing layer, as described in
\Sec\ref{subsec:magnetospheric_radius}. ST93 allow for $r_{\rm m1}$ to
decrease if the velocity shear across the mixing layer exceeds the
radial drift velocity, arguing that in this case the KHI transports
material across the boundary layer faster than it can be brought into
corotation. In our model, material is transported across the mixing
layer at the radial drift velocity. This is motivated by numerical
simulations of the KHI (\Sec\ref{sec:mhd_instabilities}) showing that
mass transport is inhibited in a thin mixing layer.

In both our model and ST93, magnetic stresses act only on the mixing
layer, so the magnitude of the magnetic torque on the disc is smaller
than in models of star-disc interactions where magnetic stresses act on
the entire disc, such as \citet{Rappaport.2004} (RFS04 henceforth). For
an order-of-magnitude comparison, one can take the representative
values $d_B = 0.01 r_{\rm c}$, $r_{\rm m1} = 1.1 r_{\rm c}$, and
integrate our assumed form of the magnetic stress $\left|S(r)\right| =
\mu^2 r^{-4} (4\pi)^{-1}$ over the mixing layer, to determine the
approximate magnitude of the magnetic torque on the mixing layer in our
model. One can similarly calculate the magnetic torque on the disc in
the RFS04 model, by integrating the assumed form of the magnetic stress
in RFS04, $\left|S(r)\right| = \mu^2 r^{-4} (2\pi)^{-1} [1-\Omega_{\rm
K}(r) / \Omega_*]$, over the entire disc outside $r_{\rm c}$. We find
that the magnetic torque on the mixing layer in our model is of order
0.03 times the magnetic torque on the disc in RFS04, assuming identical
stellar field strengths and spin frequencies in the two models. For an
alternative perspective, one can take the assumed form of the magnetic
stress in RFS04 and compute the ratio of the magnetic torque integrated
over the entire disc outside $r_{\rm c}$ to the torque integrated only
over the region $1.1 r_{\rm c} < r < 1.11 r_{\rm c}$. We find that
these torques are in the approximate ratio 1:0.008.

\subsection{Surface density of the disc}
\label{subsec:disc_surface_density}

The surface density of the disc outside $r_{\rm m2}$ evolves according
to the thin disc equation
\begin{equation} \label{eq:sigma_evolution}
   \frac{\partial \Sigma}{\partial t}
   = \frac{3}{r} \frac{\partial}{\partial r}
   \left[ r^{1/2} \frac{\partial}{\partial r}
      \left( \nu \Sigma r^{1/2} \right)
   \right],
\end{equation}
where $\nu$ is the kinematic viscosity. We neglect the possibility that
the magnetic pressure in the mixing layer causes the disc to thicken
\citep{Wang.1987}. The outer boundary condition is
\begin{equation}
   \lim_{r \rightarrow \infty} \dot{M}(r,t) = \dot{M}_{\infty},
\end{equation}
where $\dot{M}_\infty$ is some value set by the details of Roche lobe
overflow. At the inner boundary, if $r_{\rm m2}
> r_{\rm c}$, the magnetosphere presents a centrifugal barrier and one
expects $\Sigma(r_{\rm m2})$ to be larger than if $r_{\rm m2} < r_{\rm
c}$ [ST93; see also RFS04], so we choose a boundary condition of the
form $\nu \Sigma(r_{\rm m2}) \propto \exp[a(r_{\rm m2}/r_{\rm c} -
1)]$, with $a
> 0$. We fix the constant of proportionality by considering steady
state solutions, which have the form
\begin{equation} \label{eq:ss_sigma}
   \nu \Sigma =   \frac{\dot{M}}{3 \pi}
   \left[ 1 - \beta \, \left( \frac{r_{\rm m2}}{r} \right)^{1/2} \right],
\end{equation}
with $\beta < 1$ \citep{Frank.1992}. If $r_{\rm m2} = r_{\rm c}$, the
magnetosphere corotates with the mixing layer, so the disc structure
does not depend on the location of $r_{\rm m2}$ (ST93) and
(\ref{eq:ss_sigma}) implies $\beta = 0$ and $\nu \Sigma =
\dot{M}/(3\pi)$. We therefore choose the boundary condition at the
inner edge to be
\begin{equation} \label{eq:innerBC}
   \nu \Sigma(r_{\rm m2}) = \frac{\dot{M}(r_{\rm m2})}{3 \pi}
      \exp[a(r_{\rm m2}/r_{\rm c} - 1)].
\end{equation}
Comparison of (\ref{eq:ss_sigma}) and (\ref{eq:innerBC}) shows that
\begin{equation} \label{eq:beta}
   \beta = 1 - \exp[a(r_{\rm m2}/r_{\rm c} - 1)]
\end{equation}
in steady state. We set $a = 1$ in this paper, except where specified
otherwise. We show in \Sec\ref{sec:bistability} and
\Sec\ref{sec:flaw_in_framework} that our results remain qualitatively
unchanged for $0 \la a \la 2$. The boundary condition
(\ref{eq:innerBC}) differs slightly from that in ST93, where $\nu
\Sigma(r_{\rm m2}) \propto \exp[a((r_{\rm m2}/r_{\rm c})^{3/2}-1)]$ and
there is also a power law dependence of $\nu \Sigma(r_{\rm m2})$ on the
mass accretion rate.

The viscosity $\nu$ in ST93 is constant in time and decreases as a
power law in radius. In our model, we assume that $\nu$ is constant in
time and independent of radius for simplicity. Later
(\Sec\ref{subsec:radial_viscosity_gradient}), we show that including a
power law dependence of viscosity on radius leaves our results
qualitatively unchanged.

\subsection{Stellar spin frequency}
\label{subsec:stellar_ang_freq}

The stellar spin frequency, $\Omega_*$, evolves according to the
equation of motion
\begin{equation} \label{eq:omegadot}
   I \dot{\Omega}_* = N,
\end{equation}
where $I$ is the star's moment of inertia and
\begin{equation} \label{eq:torque}
   N = \dot{M}' \Omega_{\rm K}(r_{\rm m2}) r_{\rm m2}^2
      + N_\nu (r_{\rm m2})
\end{equation}
is the torque on the star, set equal to the rate of change of angular
momentum inside the surface $r = r_{\rm m2}$. There is no magnetic term
in (\ref{eq:torque}) since, by assumption, there is no magnetic stress
on the disc for $r > r_{\rm m2}$. The first term in (\ref{eq:torque})
is the material torque and the second term is the viscous torque, given
by
\begin{equation} \label{eq:viscoustorque}
   N_\nu = 2 \pi r \nu \Sigma r^2 \frac{\partial \Omega}{\partial r}.
\end{equation}
In the steady state, (\ref{eq:torque}) reduces to
\begin{equation} \label{eq:sstorque}
   N = \beta \dot{M} \Omega_{\rm K}(r_{\rm m2}) r_{\rm m2}^2.
\end{equation}
From (\ref{eq:beta}) and (\ref{eq:sstorque}), the steady state torque
on the star is positive (negative) when $r_{\rm m2}$ is less than
(greater than) $r_{\rm c}$. Away from the steady state, this is no
longer necessarily true, since (\ref{eq:beta}) and (\ref{eq:sstorque})
hold true only in the steady state. [As matter accretes onto the star,
the star's moment of inertia changes slightly, but this is a second
order effect; \citet{Ghosh.1977}.] We show that long-term changes in
$\Omega_*$ alter the equilibria of the star-disc system
(\Sec\ref{subsec:longterm_time_evol}); in contrast, ST93 held
$\Omega_*$ constant.

\subsection{Order-of-magnitude estimates of physical quantities}
\label{subsec:order_of_mag}

In this section, we estimate the key physical parameters in the problem
for typical systems. Let $M_{1.4}$, $I_{45}$, $P_{10}$, and
$\dot{M}_{-9}$ denote the mass, moment of inertia, spin period and
accretion rate of a neutron star, normalised by $M = 1.4 {~\rm
M_\odot}$, $I = 10^{45} {~\rm g\,cm^2}$, $P_{\rm spin} = 10 {~\rm s}$,
and $\dot{M} = 10^{-9} {~\rm M_\odot\,yr^{-1}}$ respectively.

The corotation radius, defined by $\Omega_* = \Omega_{\rm K}(r_{\rm
c})$, is given by
\begin{equation}
   r_{\rm c} = 7.78 \times 10^8 M_{1.4} ^ {1/3} P_{10} ^ {2/3} {\rm cm}.
\end{equation}
The characteristic torque $N_{\rm char}$ at the corotation radius is
\begin{eqnarray} \label{eq:Nchar}
   N_{\rm char} &=& \dot{M} \left({\rm G} M r_{\rm c}\right)^{1/2} \\
   &=& 2.40 \times 10^{34} \dot{M}_{-9} M_{1.4}^{2/3} P_{10}^{1/3}
   {~\rm g\,cm^2\,s^{-2}}.
\end{eqnarray}
The corresponding rate of change of the spin frequency is
\begin{equation} \label{eq:charspinup}
\dot{\Omega}_{*,{\rm char}}/(2\pi)
   = 38.2 \times 10^{-13}
   I_{45}^{-1} \dot{M}_{-9} M_{1.4}^{2/3} P_{10}^{1/3}
   {~\rm Hz\,s^{-1}}.
\end{equation}
Pulsars exhibiting torque reversals spin up (or down) at a comparable
rate, with $\left| \dot{\Omega}_* \right|/(2\pi) =
7$--$60{\rm~Hz~s}^{-1}$ \citep{Nelson.1997}. In
\Sec\ref{subsec:other_equilibria}, we show that our model predicts spin
up/down rates of this order of magnitude.

The kinematic viscosity in a disc is conventionally parametrized by the
$\alpha$ parameter \citep{Shakura.1973}. From the standard unmagnetised
$\alpha$-disc solution \citep{Frank.1992},
\begin{equation}
   \nu =
      4.0 \times 10^{12}
      \alpha^{4/5}
      \dot{M}_{-9}^{3/10}
      M_{1.4}^{-1/4}
      (r / 10^9 {~\rm cm})
      {~\rm cm^2\,s^{-1}},
\end{equation}
taking $r \approx r_{\rm c}$ and $0.001 < \alpha < 1.0$, we conclude
that $\nu$ lies in the range $10^{10}$ to $10^{13} {~\rm
cm^2\,s^{-1}}$. It is convenient to define the viscous time-scale
corresponding to the length-scale $r_{\rm c}$,
\begin{eqnarray}
   \label{eq:taunu}
   \tau_\nu &=& 2r_{\rm c}^2/(3\nu) \\
   &=& 4.06 \times 10^4 M_{1.4}^{2/3} P_{10}^{4/3} \nu_{13}^{-1} \, {\rm s},
\end{eqnarray}
with $\nu_{13} = \nu / (10^{13} {~\rm cm^2\,s^{-1}})$. Using the above
estimates for $\nu$ and $r_{\rm c}$, we find $\tau_\nu$ in the range
$1$ to $10^3 {~\rm day}$; this is the time-scale for torque transitions
in our model (\Sec\ref{subsec:short_term_time_evol}). By comparison,
\citet{Torkelsson.1998} found $\tau_\nu$ to be typically less than $0.5
{~\rm day}$

If the magnetic diffusivity is due to the same turbulence that is
responsible for the kinematic viscosity [e.g. the magneto-rotational
instability of \citet{Balbus.1991}], one expects the magnetic Prandtl
number to satisfy ${\rm Pm} = \nu_B/\nu \sim 1$. By contrast, in the
literature dealing with the diffusion of magnetic fields through discs,
${\rm Pm}$ is frequently considered as a free parameter in the range
$10^{-2} \la {\rm Pm} \la 10^2$ \citep{Reyes.1996, Lubow.1994}. The
work in this paper is valid for ${\rm Pm} \ll 1$, since for higher
${\rm Pm}$ the mixing layer is not thin (\Sec\ref{sec:bistability}).

\section{Torque Bistability}
\label{sec:bistability}

In this section, we show that the model of the disc-magnetosphere
interaction presented in \Sec\ref{sec:newmodel} possesses two stable
equilibria, with opposite signs of torque, under certain conditions. In
searching for equilibria, we note that the time-scale
$\Omega_*/\dot{\Omega}_* \sim 10^{13} {\rm~s}$ is  longer than the
time-scale $\tau_\nu$ for $r_{\rm m1}$ and $d_B$ to change. Referring
to (\ref{eq:rm1dot}) and (\ref{eq:dBdot}), we observe that the
condition for equilibrium is
\begin{equation} \label{eq:eqm_condition}
   d_B =
   \nu_B / \left| v_r(r_{\rm m2}) \right|
   = d_{\rm c}
\end{equation}

First, we obtain dimensionless expressions for $d_B$, $d_{\rm c}$ and
$r_{\rm m1,max}$ in equilibrium. From (\ref{eq:eqm_condition}),
(\ref{eq:innerBC}), (\ref{eq:mdot}), and the definition of the Prandtl
number ${\rm Pm} = \nu_B/\nu$, we find
\begin{equation} \label{eq:dB_on_rc}
   d_B/r_{\rm c} = \frac{2}{3} \frac{r_{\rm m2}}{r_{\rm c}}\, {\rm Pm} \,
      \exp [a (r_{\rm m2}/r_{\rm c} - 1)].
\end{equation}
A thin mixing layer ($d_B \ll r_{\rm c}$) therefore requires a small
Prandtl number (\Sec\ref{subsec:order_of_mag}). From (\ref{eq:dcfirst})
and (\ref{eq:magstress_spruit}), we find
\begin{eqnarray} \label{eq:dC_on_rc}
   \nonumber
   d_{\rm c} / r_{\rm c} &=&
      \xi^{7/2} \sqrt{2} \,
      (r_{\rm m1} / r_{\rm c})^6 \,
      {\rm sgn} (r_{\rm c} - r_{\rm m1}) \\
      && \times
      \left[
         \left( \frac{r_{\rm m2}}{r_{\rm m1}} \right)^2
         \left( \frac{r_{\rm m2}}{r_{\rm c}} \right)^{-3/2}
         -1
      \right],
\end{eqnarray}
where
\begin{equation} \label{eq:xi}
   \xi = \eta^{-2/7} r_{\rm c} / r_{\rm A}
\end{equation}
is a dimensionless parameter, and
\begin{equation} \label{eq:alfven_radius}
   r_{\rm A} =
      \left(
         \frac{
            \mu^4
         }{
            2 G M \dot{M}^2
         }
      \right)^{1/7}
\end{equation}
is the characteristic Alfv\'{e}n radius for spherical accretion
\citep{Elsner.1977}. From (\ref{eq:rm1max}),
(\ref{eq:magstress_spruit}) and (\ref{eq:alfven_radius}), we find
\begin{equation} \label{eq:rm1max_on_rc}
   r_{\rm m1,max} / r_{\rm c} = 2^{-3/10} \xi ^{-7/10}.
\end{equation}
Anticipating that $r_{\rm m1} \sim r_{\rm c}$, we initially restrict
our investigation to the regime $\xi \leq 2^{-3/10}$, $r_{\rm m1,max}
\geq r_{\rm c}$, so that the constraint $r_{\rm m1} \leq r_{\rm
m1,max}$ is satisfied.

We now find equilibria by plotting $d_B/r_{\rm c}$ and $d_{\rm
c}/r_{\rm c}$ against $r_{\rm m2}/r_{\rm c}$ and solving
(\ref{eq:eqm_condition}) graphically. Equations
(\ref{eq:eqm_condition}), (\ref{eq:dB_on_rc}) and (\ref{eq:dC_on_rc})
\emph{always admit one solution} for $r_{\rm m2} > r_{\rm c}$, which we
denote $A$ (see Fig.~\ref{fig:monoandbistable}). This is a spin-down
equilibrium, since the torque is negative whenever $r_{\rm m2}>r_{\rm
c}$ (\Sec\ref{subsec:stellar_ang_freq}).

\subsection{Multiple equilibria}
\label{subsec:other_equilibria}

\begin{figure}
   \centering
   \includegraphics[width=0.9\linewidth,height=!]{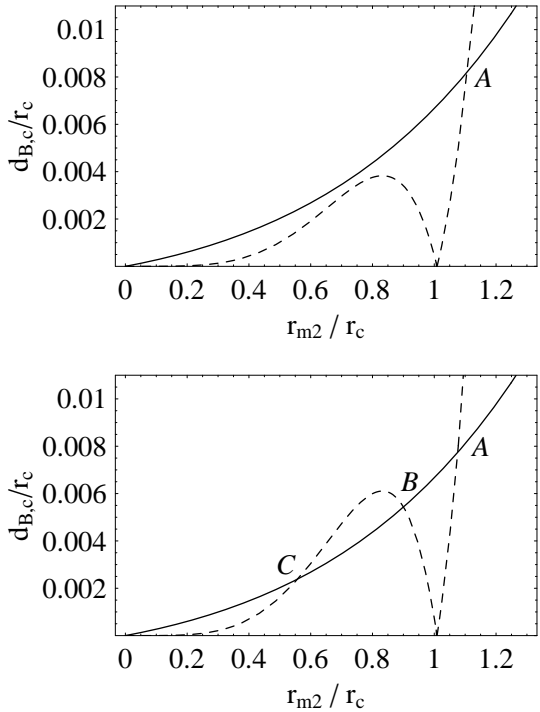}
   \caption{
      Equilibrium states for the disc-magnetosphere system. The top
      panel shows $d_B/r_{\rm c}$ (solid curve) and $d_{\rm c}/r_{\rm
      c}$ (dashed curve) as functions of $r_{\rm m2}/r_{\rm c}$, for
      ${\rm Pm} = 0.01$, $\xi = 0.35$. There is only one equilibrium,
      marked $A$, corresponding to spin down. The bottom panel shows
      the same information but for ${\rm Pm} = 0.01$, $\xi = 0.4$.
      There are two stable equilibria, marked $A$ and $C$, which
      correspond to spin-down and spin-up respectively, and one
      unstable spin-down equilibrium, marked $B$.
   }
   \label{fig:monoandbistable}
\end{figure}

\begin{figure}
   \centering
   \includegraphics[width=0.9\linewidth,height=!]{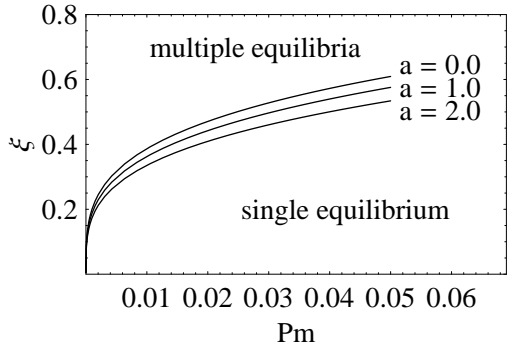}
   \caption{
      Combinations of Prandtl number ${\rm Pm}$ and ratio $\xi$ of
      corotation radius to Alfv\'{e}n radius, for which the
      disc-magnetosphere system has a single equilibrium and multiple
      equilibria. Curves for $0 \leq a \leq 2$ are plotted.
   }
   \label{fig:stability_regions}
\end{figure}

Under certain conditions, two other equilibria become available to the
system. The top panel of Fig.~\ref{fig:monoandbistable} shows that for
$\xi = 0.35$, ${\rm Pm} = 0.01$, the curves of $d_B/r_{\rm c}$ and
$d_{\rm c}/r_{\rm c}$ as functions of $r_{\rm m2} / r_{\rm c}$ have
only a single intersection, so the disc-magnetosphere system has only a
single, spin-down, equilibrium, denoted $A$. However, setting $\xi =
0.40$, ${\rm Pm} = 0.01$, as in the bottom panel of
Fig.~\ref{fig:monoandbistable}, one finds that the curves of
$d_B/r_{\rm c}$ and $d_{\rm c}/r_{\rm c}$ have three intersections, and
hence the disc-magnetosphere system has three equilibria, denoted $A$,
$B$, $C$. These equilibria have $r_{\rm m2}^{(A)} = 1.08 r_{\rm c}$,
$r_{\rm m2}^{(B)} = 0.90 r_{\rm c}$,  and $r_{\rm m2}^{(C)} = 0.56
r_{\rm c}$, so $A$ is a spin-down equilibrium while $B$ and $C$ are
spin-up equilibria. Combining (\ref{eq:sstorque}) and (\ref{eq:Nchar}),
one obtains
\begin{equation} \label{eq:Omegadotchar}
   \dot{\Omega}_* =
      \beta \,
      (r_{\rm m2} / r_{\rm c})^{1/2} \,
      \dot{\Omega}_{*,{\rm char}},
\end{equation}
giving $\dot{\Omega}_*^{(A)} = -0.082 \, \dot{\Omega}_{*,{\rm char}}$,
$\dot{\Omega}_*^{(B)} = 0.089 \, \dot{\Omega}_{*,{\rm char}}$, and
$\dot{\Omega}_*^{(C)} = 0.27 \, \dot{\Omega}_{*,{\rm char}}$. In
\Sec\ref{subsec:stability} we show that equilibria $A$ and $C$ are
stable to perturbations in $r_{\rm m1}$ and $r_{\rm m2}$, while
equilibrium $B$ is unstable. The system is thus \emph{bistable}, with
the two stable equilibria having opposite signs of torque.

A single equilibrium trifurcates into three equilibria when the curves
of $d_B/r_{\rm c}$ and $d_{\rm c}/r_{\rm c}$ in
Fig.~\ref{fig:monoandbistable} are tangential at some $r_{\rm m2} <
r_{\rm c}$. Solving for this condition numerically, we obtain
Fig.~\ref{fig:stability_regions}, which shows the combinations of ${\rm
Pm}$ and $\xi$ for which multiple equilibria exist. For $a = 1$, the
critical value $\xi_{\rm crit}$ separating the single and multiple
equilibria is well approximated by
\begin{equation} \label{eq:xicrit}
   \xi_{\rm crit} = 1.36 {\rm Pm}^{0.287},
\end{equation}
with a fractional error of $<1$\% in the range $0 < {\rm Pm} < 0.05$.

The existence of multiple equilibria can be understood as follows. One
can think of $d_{\rm c}$ as measuring how `difficult' it is for
magnetic stresses to change the angular velocity of infalling material
from $\Omega_{\rm K}(r_{\rm m2})$ to $\Omega_*$ as it crosses the
mixing layer. At $r_{\rm m2} \approx r_{\rm c}$, there is no velocity
shear across the mixing layer, so $d_{\rm c} \approx 0$. As $r_{\rm
m2}$ is displaced from $r_{\rm c}$, the velocity shear increases and
hence $d_{\rm c}$ increases. For small enough $r_{\rm m2}$, $d_{\rm c}$
decreases again because the magnetic stress is proportional to $r_{\rm
m1}^{-6}$, so it becomes `easy' for the magnetic field to bring
material crossing the mixing layer into corotation. The shape of
$d_{\rm c}$ as a function of $r_{\rm m1}$ suggests that even if $d_B$
was determined differently than in \Sec\ref{subsec:blthickness} (e.g.
with a more sophisticated model of diffusion, or modified by the action
of the KHI), $d_B(r_{\rm m2})$ and $d_{\rm c}(r_{\rm m2})$ might still
have multiple intersections and bistability might still occur.

\subsection{Stability}
\label{subsec:stability}

Suppose that an equilibrium is perturbed by slightly increasing $r_{\rm
m1}$. From (\ref{eq:rm1dot}), if $d_B - d_{\rm c} > 0$, $r_{\rm m1}$
will continue to increase, whereas if $d_B - d_{\rm c} < 0$, $r_{\rm
m1}$ will tend back to the equilibrium value. This suggests that those
equilibria for which $d_B - d_{\rm c}$ is a decreasing (increasing)
function of $r_{\rm m2}$ are stable (unstable). Thus we expect
equilibria $A$ and $C$ to be stable, and equilibrium $B$ to be
unstable.

Elaborating on the above argument, we calculate the evolution of a
perturbation about each equilibrium. As a (rough) first approximation
to the equations of motion, we assume that changes in the surface
density profile of the disc occur rapidly compared to changes in
$r_{\rm m1}$, so that $\Sigma(r,t)$ can be regarded as passing through
a succession of equilibrium states given by (\ref{eq:ss_sigma}). In
reality, $r_{\rm m1}$ and $\Sigma$ change on similar time-scales;
$\Sigma$ adjusts  over a length-scale $r_{\rm m1}$ in a time $\sim
r_{\rm m1}^2/\nu$ \citep[p.70]{Frank.1992}, while the radial drift
velocity in a thin disc
\begin{equation} \label{eq:vr}
   v_r = \frac{-3}{\Sigma r^{1/2}} \frac{\partial}{\partial r}
   \left( \nu \Sigma r^{1/2} \right),
\end{equation}
together with (\ref{eq:rm1dot}), implies $r_{\rm m1}/\dot{r}_{\rm m1}
\sim r_{\rm m1}^2/\nu$. None the less, in our approximation, we find
the following evolution equations
\begin{eqnarray} \label{eq:rm1dot_dimless}
   \tau_\nu \frac{\rm d}{{\rm d}t}\left(\frac{r_{\rm m1}}{r_{\rm c}}\right)
   &=&\frac{d_B-d_{\rm c}}{d_B}\frac{r_{\rm c}}{r_{\rm m2}}
   \exp[a(1-r_{\rm m2}/r_{\rm c})] \\
   \nonumber
   &\equiv& f_1,
\end{eqnarray}
\begin{eqnarray} \label{eq:rm2dot_dimless}
   \tau_\nu \frac{\rm d}{{\rm d}t} \left( \frac{r_{\rm m2}}{r_{\rm c}} \right)
   &=&
      \frac{2}{3} \, {\rm Pm} \, \frac{r_{\rm c}}{d_B}
      - \frac{r_{\rm c}}{r_{\rm m2}}
      \exp[a(1-r_{\rm m2} / r_{\rm c})] \\
   \nonumber
   &\equiv& f_2,
\end{eqnarray}
where the instantaneous value of $d_{\rm c}$ is given by
(\ref{eq:dC_on_rc}), $\tau_\nu$ is given by (\ref{eq:taunu}), and $d_B
= r_{\rm m2} - r_{\rm m1}$.

Linearising about an equilibrium [$r_{\rm m1}^{\rm (eq)}$, $r_{\rm
m2}^{\rm (eq)}$], we write the equations of motion in the form
$\tau_\nu \dot{x}_i = A_{ij} x_j$, with $\bmath{x} = [(r_{\rm
m1}-r_{\rm m1}^{\rm (eq)})/r_{\rm c}, (r_{\rm m2}-r_{\rm m2}^{\rm
(eq)})/r_{\rm c}]$ and $A_{ij} = \partial f_i / \partial x_j$. The
eigenvalues of $A_{ij}$ give the growth rate of perturbations.
Obtaining analytic expressions for the growth rates is not tractable,
so we calculate them numerically. For equilibria $A$, $B$ and $C$ of
Fig.~\ref{fig:monoandbistable}, we find that the dominant modes of a
perturbation have growth rates of -7.1, 3.8, and -4.7 respectively, in
units of $\tau_\nu^{-1}$, indicating that $A$ and $C$ are stable, while
$B$ is unstable.

We emphasise that the above analysis provides an indication of
stability only, since it is based on the assumption that changes in the
surface density profile of the disc are rapid compared to changes in
the position of the magnetosphere. To prove stability rigorously, one
must to solve the disc PDEs subject to our chosen boundary conditions
(e.g. ST93). However, this lies beyond the scope of this paper,
especially in view of the uncertainties surrounding the boundary
conditions themselves (e.g. the true, time dependent nature of the
magnetic field in the vicinity of the mixing layer is unknown).

\subsection{Externally driven evolution}
\label{subsec:short_term_time_evol}

\begin{figure}
   \centering
   \includegraphics[width=0.9\linewidth,height=!]{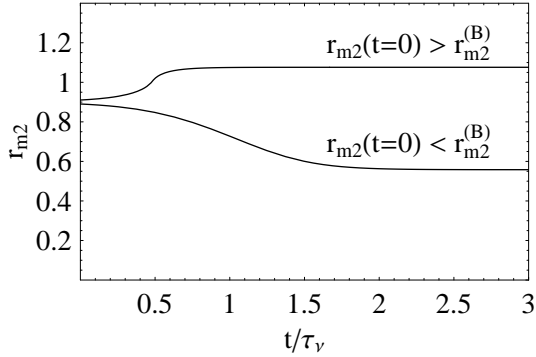}
   \caption{
      Transitions between equilibria, for the system in the bottom
      panel of Fig.~\ref{fig:monoandbistable}. If the system starts
      with $r_{\rm m2}$ larger (smaller) than $r_{\rm m2}^{(B)}$, it
      switches to equilibrium $A$ ($C$).
   }
   \label{fig:transitions}
\end{figure}

How does the bistable system in \Sec\ref{subsec:other_equilibria}
switch from one equilibrium to another? One possibility is that an
extraneous perturbation, such as a change in $\dot{M}_\infty$, pushes
the system from $A$, through $B$ (which is unstable), to $C$, in
Fig.~\ref{fig:monoandbistable}. To explore this possibility, we
integrate the equations of motion (\ref{eq:rm1dot_dimless}) and
(\ref{eq:rm2dot_dimless}), starting near the unstable equilibrium $B$.
Typical results are shown in Fig.~\ref{fig:transitions}. One finds that
the system tends to either of the stable equilibria, depending on the
initial conditions. In each case, we set the initial value of $d_B$ to
the value corresponding to equilibrium $B$, but we set the initial
value of $r_{\rm m2}$ to be slightly larger (smaller) than $r_{\rm
m2}^{(B)}$, whereupon the system tends to $A$ ($C$). In each case, the
transition between equilibria occurs on a time-scale $\sim \tau_\nu$.

\subsection{Long-term evolution due to torque feedback}
\label{subsec:longterm_time_evol}

The system can also switch from one equilibrium to another as
$\Omega_*$ changes secularly in response to the disc torque. To explore
this possibility, let us assume that the system is initially in a state
of spin-up (equilibrium $C$). From (\ref{eq:xi}), $\xi \propto r_{\rm
c} \propto \Omega_*^{-2/3}$, so as the star spins up, $\xi$ decreases,
the system moves from the upper to the lower region of parameter space
in Fig.~\ref{fig:stability_regions}, and equilibria $B$ and $C$ vanish.
The system then occupies the remaining equilibrium, $A$, and spins
down. From (\ref{eq:rm1max_on_rc}), $r_{\rm m1,max}/r_{\rm c} \propto
\xi^{-7/10} \propto \Omega_*^{7/15}$, so as the star spins down,
$r_{\rm m1,max}/r_{\rm c}$ decreases, until the state $r_{\rm m1} =
r_{\rm m1,max}$ is reached. Ultimately, $r_{\rm m1,max}$ approaches
$r_{\rm c}$, the torque tends to zero, and the system stops spinning
down.

Due to the dependences of $\xi$ and $r_{\rm m1,max} / r_{\rm c}$ on
$\Omega_*$, $\Omega_*$ undergoes fractional changes of order unity in
this evolutionary scenario. For example, suppose ${\rm Pm} = 1$, $\xi =
0.4$, and initially $\Omega_* = 1$ in arbitrary units. Then the star
spins up to $\Omega_* = 1.16$, and subsequently spins down, with
$\Omega_* \rightarrow 0.40$ as $t \rightarrow \infty$.

\subsection{Radial viscosity gradient}
\label{subsec:radial_viscosity_gradient}

\begin{figure}
   \centering
   \includegraphics[width=0.9\linewidth,height=!]{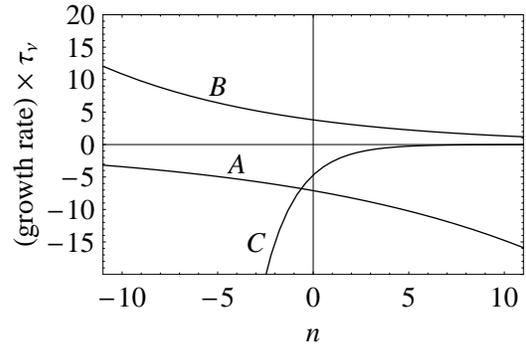}
   \caption{
      Growth rates of perturbations about equilibria $A$, $B$ and $C$,
      for the system in the bottom panel of
      Fig.~\ref{fig:monoandbistable}, as functions of the power law $n$
      governing the disc viscosity. As $n$ varies, the growth rates
      change in magnitude but not in sign.
   }
   \label{fig:growthrates}
\end{figure}

Thus far we have assumed that the viscosity is constant in time and
independent of radius, whereas in ST93 $\nu$ was allowed to vary as a
power law in radius:
\begin{equation}
   \nu(r) = \nu (r_{\rm c}) \, (r / r_{\rm c})^n.
\end{equation}
Adopting this prescription in our model, we find that the locations of
the equilibria presented in \Sec\ref{subsec:other_equilibria} are
unchanged, provided that we take ${\rm Pm}$ to be uniform (for
simplicity). This is because the derivations of (\ref{eq:dB_on_rc}) and
(\ref{eq:dC_on_rc}) do not rely upon the assumption of uniform $\nu$.
With regard to the stability analysis in \Sec\ref{subsec:stability},
one finds that the evolution equations (\ref{eq:rm1dot_dimless}) and
(\ref{eq:rm2dot_dimless}) are each modified by a factor of $(r_{\rm m2}
/ r_{\rm c})^n$ on the right hand sides, and the $\nu$ appearing in
(\ref{eq:taunu}) is given by $\nu(r_{\rm c})$. Numerical evaluation of
the dominant linear growth rate (maximum eigenvalue of $A_{ij}$), for a
range of $n$, reveals that the growth rate changes in magnitude but not
in sign (Figure~\ref{fig:growthrates}), so the stability of each
equilibrium is qualitatively unchanged from the case of uniform
viscosity.

More realistically, one might expect the viscosity $\nu$ to vary as a
function of the disc temperature and density, and therefore ${\rm Pm}$
may not be uniform. The detailed study of this scenario is beyond the
scope of this paper.

\subsection{Wind torque}

For the low magnetic Prandtl numbers adopted in this study (${\rm Pm}
\approx 10^{-2}$), significant bending of the magnetic field takes
place, which might lead to the formation of a centrifugal wind
\citep[e.g.][]{Lubow.1994,Reyes.1996}, accompanied by a magnetic torque
that removes angular momentum from the disc.

We can investigate the effect of the wind torque on our calculations
with a toy model, in which we assume that a fraction $\delta$ of the
field lines threading the mixing layer are open and form a wind, and
the magnetic stresses associated with these field lines remove angular
momentum from the mixing layer (wind torque). The remaining fraction
($1-\delta$) of field lines threading the mixing layer are closed
(connected to the star) and can either add or subtract angular momentum
from the mixing layer, as in \Sec\ref{subsec:magnetospheric_radius},
depending on the relative velocities of the star and disc. For
simplicity, we assume that the toroidal field associated with the open
field lines at the disc surface is equal in magnitude to the toroidal
field associated with the closed field lines, though its direction may
be different, giving $\left| S_{\rm wind} \right| \approx \eta \mu^2
r^{-6} (4\pi)^{-1}$.

In this crude picture, the wind torque modifies the equation for
$d_{\rm c}$, since $d_{\rm c}$ is calculated by equating the magnetic
torque on the mixing layer to the rate of change of angular momentum of
material crossing the mixing layer. In the case $r_{\rm m1} < r_{\rm
c}$, both the open and closed field lines subtract angular momentum
from the mixing layer, and one finds that equation (\ref{eq:dC_on_rc})
for $d_{\rm c}$ is unchanged. In the case $r_{\rm m1} > r_{\rm c}$, the
closed field lines add angular momentum to the mixing layer while the
open field lines subtract angular momentum from the mixing layer, and
one finds that (\ref{eq:dC_on_rc}) is modified by a factor of
$1/(1-2\delta)$ on the right hand side. For $0 < \delta < 1/2$, this
has the effect of elevating the section of the curve of $d_{\rm c}$ to
the right of $r_{\rm c}$ in Figure~\ref{fig:monoandbistable}, so that
equilibrium $A$ occurs closer to corotation. For $\delta > 1/2$, one
has $d_{\rm c} < 0$ for $r_{\rm m1} > r_c$, so equilibrium $A$
vanishes, and it is possible that there are no solutions for
equilibria; the toy model breaks down.

In our calculations, with or without the toy model for the wind, we
have neglected magnetic torques on the disc outside the mixing layer.
It is possible that there is a wind and an associated wind torque in
this region. The wind torque might even dominate the viscous torque
(see for example Pelletier 1992), and the omission of this torque is a
limitation of our model.

\section{Viscous torque at the mixing layer}
\label{sec:flaw_in_framework}

The net torque exerted on the mixing layer induces the corotation of
inflowing material and hence governs the evolution of the
magnetosphere. It is the sum of magnetic and viscous components. The
magnetic component results from the magnetic stresses integrated over
the top and bottom surfaces of the mixing layer (at $\left| z \right| =
h$), while the viscous component results from the mechanical shear
stress integrated over the surface $r = r_{\rm m2}$. In this section,
we show that the viscous torque, neglected by ST93, is actually $\sim
r_{\rm m2} / d_B$ times larger than the magnetic torque. When the
viscous torque is included in our model, only one stable equilibrium is
available to the disc-magnetosphere system under a wider range of
conditions than otherwise, although this depends sensitively on the
velocity profile near $r_{\rm m2}$.

\subsection{Ratio of viscous and magnetic torques}
\label{subsec:vistorque}

Material within $r_{\rm m1}$ corotates with the star, so the viscous
stress vanishes at the inner edge of the mixing layer. By assumption,
the angular velocity profile of the disc is approximately Keplerian
beyond $r_{\rm m2}$. The viscous torque $N_{\nu}$ acting on the mixing
layer is therefore
\begin{eqnarray} \label{eq:Nvis}
   N_{\nu} &=&
   2 \pi r_{\rm m2} \nu \Sigma r_{\rm m2}^2
   \left(\frac{\partial \Omega}{\partial r} \right)_{r = r_{\rm m2}} \\
   \label{eq:NvisKep}
   &=& -3 \pi \nu \Sigma(r_{\rm m2}) \, \Omega_{\rm K}(r_{\rm m2}) \, r_{\rm m2}^2.
\end{eqnarray}
The negative sign indicates that the viscous torque subtracts angular
momentum from material in the mixing layer. In the steady state, at
$r_{\rm m2}$, the surface density is given by $\nu \Sigma(r_{\rm m2}) =
\dot{M}(1 - \beta)/(3\pi)$, implying
\begin{equation} \label{eq:Nvis2}
   \left| N_{\nu} \right| = \dot{M} \Omega_{\rm K}(r_{\rm m2}) \, r_{\rm m2}^2
   \left| 1 - \beta \right|.
\end{equation}
Note that, for fixed $\dot{M}$, $\left| N_{\nu} \right|$ does not tend
to zero in the limit $\nu \rightarrow 0$, since (\ref{eq:Nvis2}) is
independent of $\nu$. In this limit, from (\ref{eq:vr}), $v_r \propto
\nu$ tends to zero, and $\Sigma \propto \dot{M} / v_r \propto \nu^{-1}$
grows without bound, in such a way that $N_\nu \propto \nu \Sigma$
remains fixed.

The magnetic torque $N_B$ on the mixing layer is
\begin{equation} \label{eq:Nmag}
   N_B = -2 \pi r_{\rm m1} d_B S(r_{\rm m1}) \, r_{\rm m1}.
\end{equation}
From (\ref{eq:rm1max}), with $r_{\rm m1} \approx r_{\rm m1,max}$ for
order-of-magnitude purposes, we obtain $\dot{M} \Omega_* \approx \pi
r_{\rm m1} \left| S(r_{\rm m1}) \right|$, implying
\begin{equation} \label{eq:Nmag2}
   \left| N_B \right| \approx 2 r_{\rm m1} d_B \dot{M} \Omega_*.
\end{equation}
Dividing (\ref{eq:Nvis2}) by (\ref{eq:Nmag2}), we find
\begin{eqnarray} \label{eq:Nvis_on_Nmag}
   \nonumber
   \left| \frac{N_{\nu}}{N_B} \right|
   & \approx & \frac{1}{2} \frac{\Omega_{\rm K}(r_{\rm m2})}{\Omega_*}
   \frac{r_{\rm m2}^2}{r_{\rm m1} d_B} \left| 1 - \beta \right| \\
   & \approx & \frac{r_{\rm m2}}{2 d_B} \left| 1 - \beta \right|,
\end{eqnarray}
where we have used $\Omega_{\rm K} (r_{\rm m2}) \approx \Omega_*$
(close to corotation) and $r_{\rm m1} \approx r_{\rm m2}$ (thin mixing
layer). The expression (\ref{eq:Nvis_on_Nmag}) shows that the viscous
torque is much greater than the magnetic torque for $d_B \ll r_{\rm
m2}$. Neglecting the viscous torque is not permissible, except for
$\beta$ close to unity.

\subsection{Equilibrium states}
\label{subsec:implications_for_our model}

\begin{figure}
   \centering
   \includegraphics[width=0.9\linewidth,height=!]{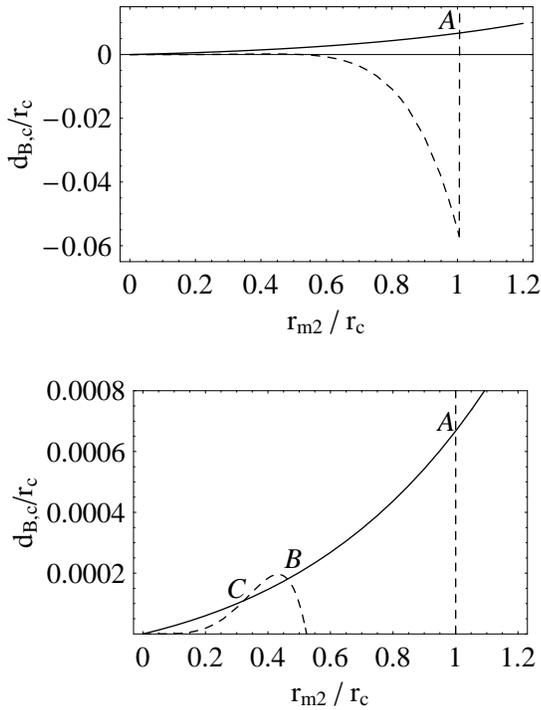}
   \caption{
      Equilibrium states for the disc-magnetosphere system including
      the viscous torque at the mixing layer. The top panel shows
      $d_B/r_{\rm c}$ (solid curve) and $d_{\rm c} / r_{\rm c}$ (dashed
      curve) as functions of $r_{\rm m2} / r_{\rm c}$, for ${\rm Pm} =
      0.01$, $\xi = 0.4$. There is only one equilibrium, marked $A$,
      with zero net torque. The bottom panel shows the same information
      but for ${\rm Pm} = 0.001$, $\xi = 0.4$. There are two stable
      equilibria, marked $A$ and $C$, corresponding to zero torque and
      spin-up respectively, and one unstable, spin-up equilibrium,
      marked $B$.
   }
   \label{fig:monoandbistablevis}
\end{figure}

\begin{figure}
   \centering
   \includegraphics[width=0.9\linewidth,height=!]{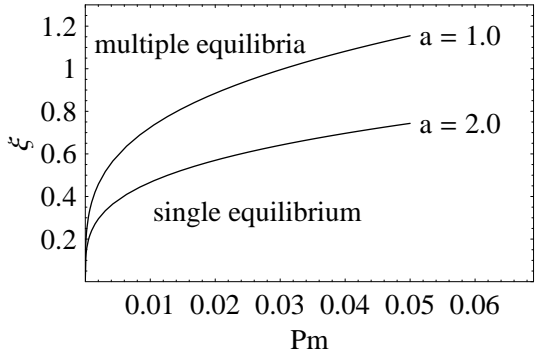}
   \caption{
      Regions in ${\rm Pm}$--$\xi$ parameter space for which the system
      has a single equilibrium or multiple equilibria, with the viscous
      torque included.
   }
   \label{fig:stabilityregionsvis}
\end{figure}

\begin{figure}
   \centering
   \includegraphics[width=0.9\linewidth,height=!]{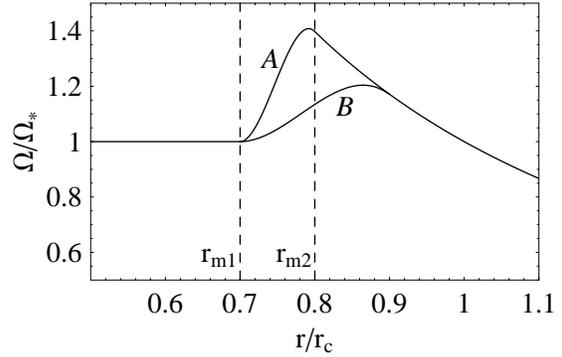}
   \caption{Schematic diagram of possible angular velocity profiles
   near the mixing layer (not to scale). Profile $A$ is Keplerian
   beyond $r_{\rm m2}$, and corresponds to a -ve torque on the boundary
   layer ($\partial \Omega/\partial r < 0$ at $r_{\rm m2}$). Profile
   $B$ is non-Keplerian for a short distance beyond $r_{\rm m2}$, and
   corresponds to a +ve torque ($\partial \Omega / \partial r > 0$ at
   $r_{\rm m2}$).
   }
   \label{fig:blcloseup}
\end{figure}

In this section, we examine the modified equilibria of the
disc-magnetosphere system after taking into account the viscous torque
(\ref{eq:Nvis}). Following a similar analysis to that presented in
\Sec\ref{subsec:other_equilibria}, one finds that equation
(\ref{eq:dB_on_rc}) for $d_B/r_{\rm c}$ is still valid, but
(\ref{eq:dC_on_rc}) needs to be modified as follows:

\begin{eqnarray} \label{eq:dC_on_rc_vis}
   \nonumber
   d_{\rm c} / r_{\rm c} &=&
      \xi^{7/2} \sqrt{2} \,
      (r_{\rm m1} / r_{\rm c})^6 \,
      {\rm sgn} (r_{\rm c} - r_{\rm m1}) \\
      && \times
      \left[
         \beta
         \left( \frac{r_{\rm m2}}{r_{\rm m1}} \right)^2
         \left( \frac{r_{\rm m2}}{r_{\rm c}} \right)^{-3/2}
         -1
      \right],
\end{eqnarray}
with $\beta$ given by (\ref{eq:beta}).

Examples of graphical solutions for equilibria, with the viscous torque
included, are shown in Fig.~\ref{fig:monoandbistablevis}. The top panel
uses the same values of dimensionless parameters (${\rm Pm} =0.01$,
$\xi = 0.4$) as the bottom panel of Fig.~\ref{fig:monoandbistable}, but
clearly the shape of $d_{\rm c}$ as a function of $r_{\rm m2}$ is very
different, and there is only one equilibrium. The bottom panel, with
${\rm Pm}=0.001$, $\xi = 0.4$, shows a system with three equilibria;
$A$ and $C$ are stable, with zero and spin-up torque respectively,
while equilibrium $B$ is unstable. In both panels, the discontinuity in
$d_{\rm c}$ arises because the magnetic torque changes sign at $r_{\rm
m2} = r_{\rm c}$, while the viscous torque does not; in reality, the
change in the magnetic torque occurs gradually over the domain $r_{\rm
c} - d_B < r_{\rm m2} < r_{\rm c} + d_B$.
Fig.~\ref{fig:stabilityregionsvis} shows the regions in ${\rm
Pm}$--$\xi$ parameter space for which there are multiple equilibria.
Note that, for a given $\xi$, the range of ${\rm Pm}$ giving rise to
multiple equilibria is smaller compared to the case where the viscous
torque is neglected (Fig.~\ref{fig:stability_regions}).

The viscous torque on the mixing layer depends on $\partial \Omega /
\partial r$ at $r_{\rm m2}$ through (\ref{eq:Nvis}). The derivation of
(\ref{eq:dC_on_rc_vis}) assumes that the disc velocity profile is
Keplerian beyond $r_{\rm m2}$, so that the viscous torque is given by
(\ref{eq:NvisKep}). This assumption is questionable, because viscous
stresses may result in a nonkeplerian velocity profile for a short
distance beyond $r_{\rm m2}$. Fig.~\ref{fig:blcloseup} sketches two
hypothetical velocity profiles in the vicinity of the mixing layer. In
one case, the velocity profile is Keplerian outside $r_{\rm m2}$, and
the viscous torque on the mixing layer is negative, while in the other
case the velocity profile is non-Keplerian for a short distance beyond
$r_{\rm m2}$, and the viscous torque on the mixing layer is positive.
Consequently, the system is very sensitive to the velocity profile near
the mixing layer. ST93 did not address this issue, because the viscous
torque on the mixing layer was neglected in their model. In the special
case $\partial \Omega / \partial r = 0$ at $r = r_{\rm m2}$, the
viscous torque on the mixing layer vanishes, and the analysis in
\Sec\ref{sec:bistability} and ST93 is valid.

\section{Turbulence in the mixing layer}
\label{sec:mhd_instabilities}

The interaction between the accretion disc and the stellar magnetic
field at the disc-magnetosphere boundary is not well understood. The
boundary is subject to magnetohydrodynamic (MHD) instabilities, e.g.
the magnetic interchange instability (MII) and the Kelvin-Helmholtz
instability (KHI), which control the turbulent transport of material
through the mixing layer, into the magnetosphere, and onto the star
\citep{Ghosh.1978, Ghosh.1979a, Ghosh.1979b, Scharlemann.1978,
Spruit.1993}. Such instabilities may also explain the quasi-periodic
oscillations in X-ray sources \citep{Baan.1979,Li.2004}.

In this section, we assess the validity of the claim that the KHI
transports material across the mixing layer with a speed (ST93)
\begin{equation} \label{eq:vKH_ST}
   v_{\rm KH} = r_{\rm m1}
   \left|\Omega_{\rm K}(r_{\rm m1}) - \Omega_*\right|.
\end{equation}
To this end, we wrote a vortex-in-cell (VIC) code to perform two
dimensional (2D), inviscid, incompressible, hydrodynamic simulations of
the KHI (with $B = 0$ and effective gravity $g_{\rm eff} = 0$) in an
elongated domain, representing a thin mixing layer (ST93). We emphasise
that the simulations give some insight into the behaviour of the KHI
but they do not faithfully simulate the global physics of the mixing
layer. Ideally, in a full treatment, one would perform simulations in
an unbounded domain and see how thick the mixing layer grows of its own
accord, rather than enforcing a thin layer and testing what sort of
flow is consistent with this assumption, as we do here.

\subsection{Vortex-in-cell code}

The VIC algorithm \citep{Liu.2000} discretises the vorticity field
$\omega = \nabla \times \bmath{v}$ into point vortices, or
\emph{vortons}. The vortons are tracked as particles, while the
velocity field $\bmath{v} = \nabla \times \bmath{A}$ is obtained by
solving Poisson's equation,
\begin{equation} \label{eq:poisson}
   \nabla^2 \bmath{A} = - \bmath{\omega},
\end{equation}
on a grid, where $\bmath{A}$ is the vector potential for the velocity
field. Vortons are advected by the velocity field; in turn, one can
reconstruct the velocity field from the vorton distribution.

Our simulations are performed in a 2D box of dimensions $0 < x < W =
1$, $-L/2 < y < L/2$. The velocity field is tracked on a regular
cartesian grid $(x,y)$ tiled by $n_x \times n_y$ cells. The domain is
periodic in one direction,
\begin{equation}
   \bmath{v}(x=0,y,t)=\bmath{v}(x=1,y,t),
\end{equation}
consistent with the cylindrical geometry of the  annular mixing layer.
In the other direction, representing the inner and outer edges of the
annulus, we set
\begin{equation}
   v_y(x,-L/2,t) = v_y(x, L/2,t) = 0,
\end{equation}
\begin{equation}
   \bomega(x,-L/2,t) = \bomega(x,L/2,t) = 0.
\end{equation}
To be absolutely consistent with ST93, one would prefer to specify
$v_x$ at $\left| y \right| = L/2$. Doing so, however, requires the
creation of new vortons at these boundaries, a feature that is hard to
implement in our code. Initially, the fluids in the inner and outer
halves of the annulus are approximately counter-streaming, with
$\bmath{v}(x,y,0) = 0.5 \bmath{\hat{x}}$ for $y < 0$ and
$\bmath{v}(x,y,0) = -0.5 \bmath{\hat{x}}$ for $y > 0$. Thus, one unit
of time in our simulations is equal to the length of the simulation
domain divided by the shear speed; physically, this is approximately
the spin period of the neutron star. To provide an initial perturbation
from which the KHI can grow, each vorton is given a random initial $y$
coordinate in the range $-0.005 < y < 0.005$, as in the top panel of
Fig.~\ref{fig:supers}.

\subsection{Transport speed}
\begin{figure}
   \centering
   \includegraphics[width=0.9\linewidth,height=!]{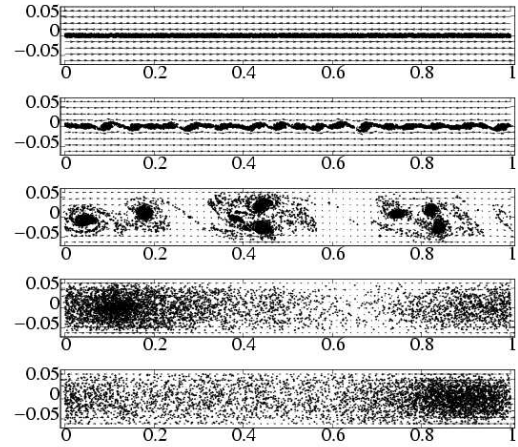}
   \caption{
      KHI instability in a domain of dimensions $W = 1.0$, $L = 1/8$,
      at times t = 0, 0.125, 1, 10, 100 (top to bottom). Diamonds
      represent vortons, and arrows show the velocity field.
   }
   \label{fig:supers}
\end{figure}

\begin{figure}
   \centering
   \includegraphics[width=0.9\linewidth,height=!]{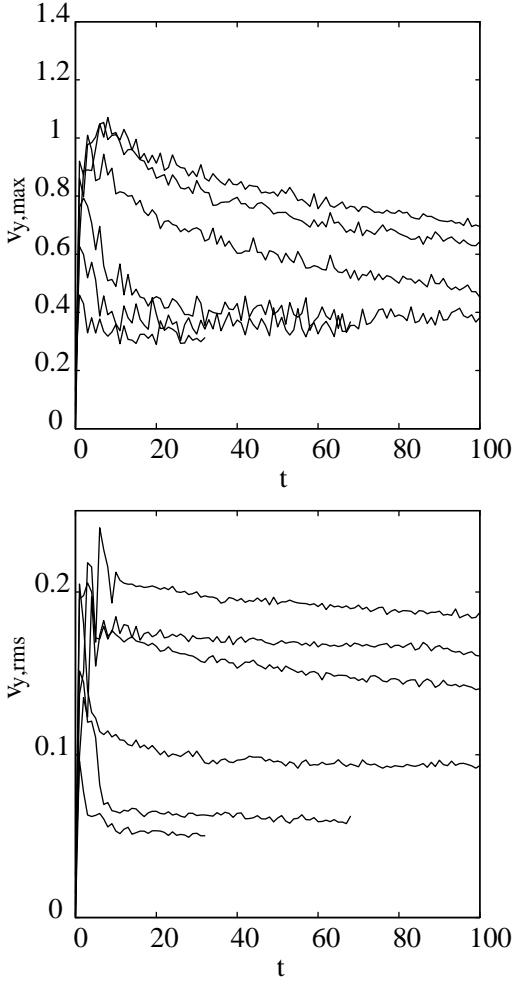}
   \caption{
      Maximum transport speed $v_{y,{\rm max}}$ (top panel) and
      root-mean-square transport speed $v_{y,{\rm rms}}$ (bottom
      panel) due to the KHI across an elongated domain, versus time.
      One unit of time is approximately one stellar spin period. In
      each panel, curves are plotted for $W/L$ = 1, 2, 4, 8, 16, 32
      (top to bottom).
   }
   \label{fig:vy_t}
\end{figure}

\begin{figure}
   \centering
   \includegraphics[width=0.9\linewidth,height=!]{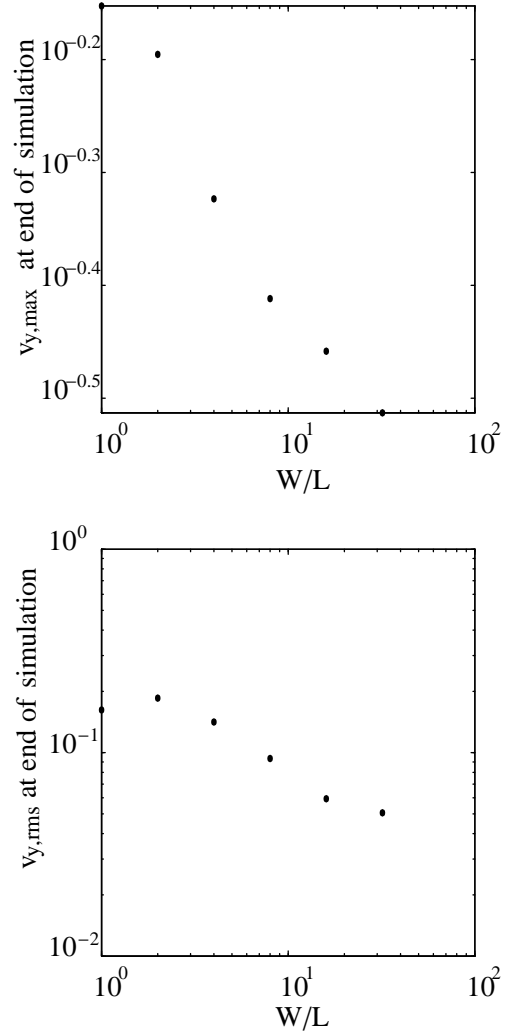}
   \caption{
      The mean of $v_{y,{\rm max}}$ (top panel) and $v_{y,{\rm rms}}$
      (bottom panel) over the last 5 time units of each simulation, as
      a function of $W/L$. One unit of time is approximately one
      stellar spin period.
   }
   \label{fig:vy_wd}
\end{figure}

The simulations track the root-mean-square and the maximum of the $y$
component of velocity, $v_{y,{\rm rms}}$ and $v_{y,{\rm max}}$, from
which we can estimate the average and maximum rates at which material
is transported across the mixing layer. We performed simulations for
$L$ = 1, 1/2, 1/4, 1/8, 1/16, 1/32, with $n_y = 64$, $n_x = 64 / L$,
and $5\times10^3$ vortons. Fig.~\ref{fig:supers} shows that the first
vortical structures form on the smallest scales $\lambda$, consistent
with the linear KHI growth rate $\gamma_{\rm KHI} \propto
\lambda^{-1}$. These structures merge hierarchically until a single
vortex occupies the simulation domain; energy cascades from small to
large scales in 2D \citep{Batchelor.1953}. The top panel of
Fig.~\ref{fig:vy_t} shows that $v_{y,{\rm max}}$ rises initially, then
decays within $\sim 20$ crossing times, asymptoting to a value that
depends upon $W/L$. The top panel of Fig.~\ref{fig:vy_wd} shows that
higher aspect ratios $W/L$ give a lower saturated value of $v_{y,{\rm
max}}$. The behaviour of $v_{y,{\rm rms}}$ as a function of $W/L$ is
similar to $v_{y,{\rm max}}$ with $v_{y,{\rm rms}} \approx 0.2
v_{y,{\rm max}}$ (bottom panels of Fig.~\ref{fig:vy_t} and
Fig.~\ref{fig:vy_wd}).

Our results imply that when the KHI is confined to an elongated domain,
the maximum transport speed across the mixing layer is significantly
less than the velocity shear. However, these results must be
interpreted with caution. Our simulations are 2D; they neglect the
Coriolis force, gravity, magnetic fields and compressibility; and the
vorticity vanishes artificially at the boundaries. One would expect the
inclusion of three-dimensionality, magnetic fields and compressibility
to decrease $v_{y,{\rm max}}$ further, since energy cascades from large
to small scales in 3D \citep{Batchelor.1953}, compressibility tends to
suppress the KHI \citep{Wang.1984}, and magnetic fields tend to disrupt
large eddies \citep{Malagoli.1996}. In contrast, it is hard to predict
the effect of different boundary conditions, the Coriolis force, and
gravity ($g_{\rm eff} \neq 0$).

\section{Discussion and Conclusions}
\label{sec:conclusions}

In this paper, we present a generalized model of the disc-magnetosphere
interaction, which incorporates diffusion of the stellar magnetic field
into the disc. We show that, for $\xi \ga {\rm Pm}^{0.3}$, the system
possesses two stable equilibria, corresponding to spin-up and
spin-down. (Our calculations indicate rather than prove stability,
since they are based on the assumption that changes in the surface
density profile of the disc are rapid compared to changes in the
position of the magnetosphere). We suggest that transitions between
stable equilibria can be induced, on the viscous time-scale $\tau_\nu$,
by changes in the mass accretion rate at the outer edge of the disc
(extraneous) or changes in the spin frequency of the star (torque
feedback). We also show that the viscous torque $N_\nu$ is generally
much larger than the magnetic torque $N_B$; when it is included, the
dynamics of the disc-magnetosphere system are very sensitive to the
velocity profile at $r_{\rm m2}$, but typically there are fewer
combinations of ${\rm Pm}$ and $\xi$ for which the system is bistable.
Preliminary numerical simulations of the KHI indicate that the
transport speed across the mixing layer between the disc and
magnetosphere is less than the shear speed when the layer is thin. In
our model, material is transported across the mixing layer at the
radial drift velocity.

Given the presence of two stable equilibria in our model, it is
interesting to ask whether we can draw any parallels with the
phenomenon of torque reversals \citep{Nelson.1997}. The X-ray pulsars
\mbox{GX 1+4}, \mbox{4U-1626-67}, \mbox{Cen X-3} and \mbox{OAO
1657-415} have all been observed to alternate between episodes of
spin-up and spin-down, such that the time-scale for transitions between
episodes (days) is much shorter than the the duration of individual
torque episodes (weeks or years). A successful model for torque
reversals should explain the observed properties: (i) Transitions never
occur between torques of the same sign (but different magnitudes). (ii)
The torque ranges from $0.2 N_{\rm char}$ for \mbox{4U-1626-67} to $2.3
N_{\rm char}$ for \mbox{Cen X-3}, assuming
$\dot{M}=10^{-9}{\rm~M_\odot~yr^{-1}}$. (iii) The transition time-scale
is much shorter than the time-scale for a sustained torque episode.
(iv) The torque and luminosity are anti-correlated in the spin-down of
\mbox{GX 1+4}. Evaluating our model against these criteria, we see the
following: (i) The two stable equilibria have opposite signs of torque,
so there are never transitions between torques of the same sign. (ii)
The magnitude of the torque is $\sim 0.1 N_{\rm char}$. (iii)
Transitions between equilibria occur over $\sim 1 {\rm~day}$
($\tau_\nu$) for large values of viscosity ($\alpha \sim 1$), as
observed, but are slow ($\sim 50 {\rm~day}$) for $\alpha \sim 0.01$
(iv) The anti-correlated torque and X-ray luminosity in \mbox{GX 1+4}
is not explained by our model. Also, our model predicts that, in the
absence of extraneous perturbations, a star that is spinning up
initially eventually transitions to spin down, then evolves toward a
state of zero torque, i.e. the model does not exhibit repeated torque
reversals in its current, simple form.

The model we have proposed, based on ST93, omits several effects which
may be important. First, the disc may generate its own magnetic field
via an MHD dynamo, driven by the shear in the mixing layer (as at the
solar tachocline). \citet{Torkelsson.1998} proposes dynamo action as a
mechanism for torque reversals. Second, material may be transported off
the disc to form an outflow or be funnelled onto the stellar poles. For
the low magnetic Prandtl numbers adopted in our study (${\rm Pm}
\approx 10^{-2}$), significant bending of the field lines takes place
which might lead to a centrifugal wind \citep{Reyes.1996,Lubow.1994}.
In the disc outside the mixing layer, the magnetic torque due to such a
wind might dominate the viscous torque
\citep{Blandford.1982,Pelletier.1992}, whereas our model neglects
magnetic torques outside of the mixing layer. Third, the mixing layer
may not be thin in the radial direction. For example, even if the
magnetic field does not penetrate far into the disc, the KHI may cause
the mixing layer to thicken. Fourth, we assume the neutron star to be
an aligned rotator, whereas the disc-magnetosphere system behaves
differently for an oblique rotator \citep{Wang.1997}. Finally, our
model does not solve for the disc structure in detail. Ideally, one
would combine a detailed solution for the disk structure
\citep[e.g.][]{Rappaport.2004} with a detailed model for the location
of the inner disk radius (which includes magnetic diffusion among other
effects).

\section*{Acknowledgments}

This research was funded in part by Australian Research Council
Discovery Project grant DP0208735.

\bibliographystyle{mn2e}
\bibliography{paper1refsv2}

\bsp

\label{lastpage}

\end{document}